\documentclass[journal=jpcbfk,manuscript=article]{achemso}
\usepackage{graphicx}
\usepackage{amssymb,amsmath}
\usepackage{textcomp}
\title{Nonequilibrium effects in DNA microarrays: a multiplatform study} 
\author{J-C. Walter}
\email{jean-charles.walter@fys.kuleuven.be}
\affiliation{Institute for Theoretical Physics, KULeuven, Celestijnenlaan 200D, B-3001 Leuven, Belgium}
\author{K. M. Kroll}
\affiliation{Institute for Theoretical Physics, KULeuven, Celestijnenlaan 200D, B-3001 Leuven, Belgium}
\author{J. Hooyberghs}
\affiliation{Institute for Theoretical Physics, KULeuven, Celestijnenlaan 200D, B-3001 Leuven, Belgium}
\alsoaffiliation{Flemish Institute for Technological Research (VITO),
Boeretang 200, B-2400 Mol, Belgium}
\alsoaffiliation{Hasselt University, Campus Diepenbeek, B-3590 Diepenbeek, Belgium}
\author{E. Carlon}
\affiliation{Institute for Theoretical Physics, KULeuven, Celestijnenlaan 200D, B-3001 Leuven, Belgium}
\begin{document}
\begin{abstract}
It has recently been shown that in some DNA microarrays the time needed to reach thermal
equilibrium may largely exceed the typical
experimental time, which is about $15h$ in standard protocols (Hooyberghs
et al. Phys. Rev. E {\bf 81}, 012901 (2010)).  In this paper we discuss
how this breakdown of thermodynamic equilibrium could be detected in
microarray experiments without resorting to real time
 hybridization data, which are difficult to implement in standard experimental
conditions. The method is based on the analysis of the distribution of
fluorescence intensities $I$ from different spots for probes carrying
base mismatches. In thermal equilibrium and at sufficiently low concentrations,
 $\log I$ is expected to be linearly related to the
hybridization free energy $\Delta G$ with a slope equal to $1/RT_{\rm
exp}$, where $T_{\rm exp}$ is the experimental temperature and $R$ is the
gas constant.  The breakdown of equilibrium results in the deviation from
this law.  A model for hybridization kinetics explaining the observed experimental
behavior is discussed, the so-called 3-state model.
 It predicts that deviations from equilibrium
 yield a proportionality of $\log I$ to $\Delta G/RT_{\rm eff}$. Here, $T_{\rm
eff}$ is an ``effective" temperature, higher than the experimental one.
This behavior is indeed observed in some experiments on Agilent arrays~\cite{hooy09,hooy10}.
We analyze experimental data from two other microarray platforms and
discuss, on the basis of the results, the attainment of equilibrium in
these cases. Interestingly, the same 3-state model predicts a (dynamical) 
saturation of the signal at values below the expected one at equilibrium.
\end{abstract}

\maketitle

\section{Introduction}

While nucleic acid hybridization has been quantitatively characterized for
strands binding in a bulk solution \cite{bloo00}, a microscopic-based
understanding of hybridization in microarrays is still lacking.  Compared
to hybridization in solution, the characterization and understanding of
physicochemical properties arising from hybridization in DNA microarrays
is more challenging, and several groups dedicated their research to this
subject~\cite{held03,heks03,zhan03,haga04,bind05b,carl06,fish07_sh,weck07,zhan07,burd08,nais08a} 
(see also the reviews in refs~\cite{halp06,levi05,bind06}).

There are two main reasons which make this task difficult. The first
one arises due to differences between different microarray platforms;
for example the fabrication method, surface chemistry, the length of
the surface-tethered probe strands and their density might significantly
differ from platform to platform. Since it is a priori rather difficult
to assess the influence of the above-mentioned factors (for a theoretical
analysis of factors influencing hybridization in microarrays see refs.~\cite{halp06} and \cite{bind06}),  
it is necessary to investigate the properties and
characteristics of each platform separately.  The second reason arises
due to the complexity of the samples (i.e., cell extract) which are used
in most hybridization experiments.  In these experiments, next to the
intended hybridization between a probe and its complementary target in
solution, a wide range of other reactions may occur: cross hybridization
between partially complementary targets in solution, target folding, and
cross hybridization between probes and partially complementary targets.
Therefore, although a large number of biological data is available,
it is difficult to disentangle from these experiments the properties
of hybridization between targets and surface-bound probes from other
spurious effects. A better understanding of the microscopic properties of
hybridization in DNA microarrays involves the analysis of well-controlled
experiments with simple target solutions.

In this context, we recall the outcome of recent
experiments~\cite{hooy10}, in which a solution containing a single target
sequence at different concentrations was hybridized to a large number of
mismatched probe sequences. It was shown that the general assumption of
thermodynamic equilibrium, commonly used when discussing experimental
data, does not hold. Apart from using a simple hybridization target
solution, these experiments were carried out at standard conditions
with respect to the buffer, the temperature, and the hybridization
times. The lack of thermal equilibrium was shown to cause a decrease
in specificity of microarrays~\cite{hooy10}, therefore it implies very
practical consequences: it limits the performance of the microarrays
below their maximal attainable level.

The aim of this paper is to discuss the microscopic origins of the
breakdown of thermal equilibrium and to investigate experiments on
different microarray platforms where such effects might be of relevance.
We start by reviewing the two-state Langmuir model of DNA hybridization
and discuss a recent extension. The latter is termed three-state
 model and includes a long-lived partially hybridized state.
  This model offers a plausible
explanation of the observed nonequilibrium behavior. We then turn our
attention to three different platforms for which controlled hybridization
experiments are available, and discuss evidence of the breakdown of
equilibrium in these cases.

\section{Two- vs. Three-State Kinetic Model}
\label{2vs3state_model}

Hybridization in solution as well as in DNA microarrays is commonly
described by a two-state process using the so-called Langmuir model
(see, e.g.,~\cite{bind06} for a review). According to the Langmuir model
the fraction of hybridized probes, $\theta$, is given by
\begin{eqnarray}
\frac{d \theta}{dt} = ck_1 (1-\theta) - k_{-1} \theta,
\label{two_states_kin}
\end{eqnarray}
where $k_1$ and $k_{-1}$ are the on- and off-rates, and $c$ is the target
concentration in solution. The rates are linked to the equilibrium free
energy $\Delta G$ via $k_1/k_{-1} =  \exp (-\Delta G/RT)$, where  $T$
is the temperature, and $R$ the gas constant.  Note that by convention
$\Delta G < 0$, and strong target-probe affinities thus imply larger
negative values for $\Delta G$.  The two-state hybridization model is
assumed to be valid if the hybridizing sequences are sufficiently short
such that intermediate states can be ignored. In solution, the zippering
of two hybridizing sequences is sufficiently rapid, and the two-state
assumption is expected to be applicable to oligomers containing up to
$30$ nucleotides~\cite{bloo00}.

By imposing $d\theta/dt=0$ on Eq.~(\plainref{two_states_kin}), we find the
equilibrium value
\begin{eqnarray}
\theta_{\rm eq} = \frac{ck_1/k_{-1} }{1+c k_1/k_{-1}},
\label{2state_eq}
\end{eqnarray}
which becomes
\begin{eqnarray}
\theta_{\rm eq} \approx c k_1/k_{-1} = c \exp (-\Delta G/RT),
\label{2state_lowc}
\end{eqnarray}
in the limit $ck_1 \ll k_{-1}$ (i.e. low concentration and/or weak binding).

\begin{figure}[t]
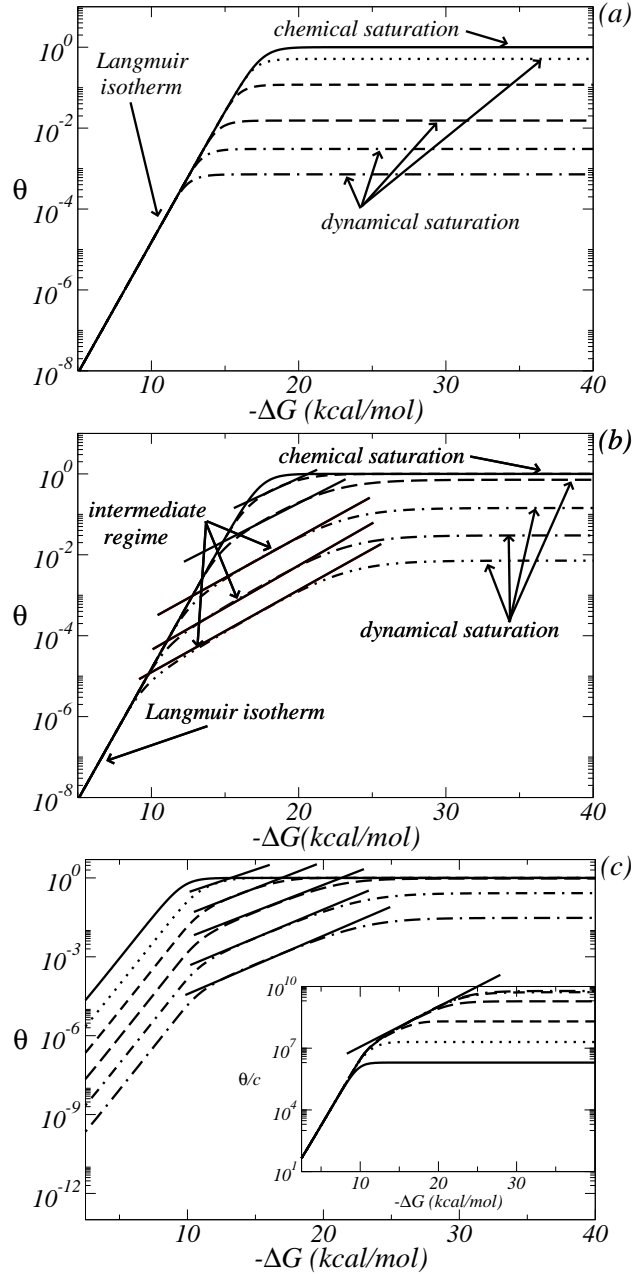

\includegraphics[width=0.5\columnwidth]{2state_model.eps}

\includegraphics[width=0.5\columnwidth]{3state_model.eps}

\includegraphics[width=0.5\columnwidth]{3state_model_tfix.eps}
\caption{Plot of $\log \theta$ vs. $\Delta G$ for the two- and three-state models. (a) Two-state model with $c=5pM$,
 $\alpha=10^4{s}^{-1}{M}^{-1}$, and $RT=0.67kcal.mol^{-1}$. The curve with solid line is the equilibrium isotherm
(Eq.~(\plainref{2state_eq})), the other lines correspond to plots of
$\theta(t)$ (Eq.~(\plainref{2state_kin})) for $t = \{4,17, 86, 700;
4000\}h$ (bottom to top).  (b) Three-state model for $c=5pM$,
$\alpha=10^5 {s}^{-1}{M}^{-1}$, $\omega=1 {s}^{-1}$, $RT=0.67kcal.mol^{-1}$,
and $\gamma=1/3$~: Sum of the solutions of Eqs.~(\plainref{dtheta1dt})
and (\plainref{dtheta2dt}) ($\theta=\theta_1+\theta_2$) vs. $\Delta G$
for different $t$ (same as in (a)) with equilibrium value of $\theta$
(solid line). An effective temperature $T_{\rm{eff}}>T_{\rm{exp}}$
can be defined for the intermediate regime (solid straight lines).
(c) Three-state model: Fixed hybridization time $t=17h$ and different
concentrations varying between $5pM$ and $500nM$ by steps of a factor
$10$ (from bottom to top). Remaining parameters identical to those in (b). System equilibrated
for the largest concentration. The intermediate regime with an effective
temperature $T_{\rm eff}$ can also be observed (solid straight lines).
Inset: vertical axis shows fraction of hybridized probes divided by the
concentration, i.e.~$\theta$/c.}
\label{fig_2and3}
\end{figure}

In this paper we consider for fixed target concentrations $c$ $\theta$ vs. $\Delta G$ plots (in this section),
 or equivalently, intensity $I$ vs. $\Delta G$ (in section describing experiments). 
Equation~(\plainref{2state_lowc}) shows that far from chemical saturation
($\theta \ll 1$) the equilibrium isotherm is characterized by the linear
relationship $\log \theta \propto \Delta G$,
with a slope equal to $1/RT$. 
The full solution of Eq.~(\plainref{two_states_kin}) is
\begin{eqnarray}
\theta (t) = \theta_{\rm eq} (1 - e^{-t/\tau}), 
\label{2state_kin}
\end{eqnarray}
where $\tau$ is the characteristic time such that
\begin{eqnarray}
\tau^{-1} = c k_1+ k_{-1}.
\label{2state_tau}
\end{eqnarray}
In order to understand the relaxation to equilibrium 
of sequences with different
$\Delta G$ , we make a hypothesis concerning the
rate constants.  Experiments for hybridization in solution and
in microarrays show that $k_1$ only weakly depends on the sequence
composition~\cite{glaz06_sh,cant80}. Typically, there is a much stronger sequence
dependence of the off-rate $k_{-1}$. We thus approximate
\begin{eqnarray}
k_1 = \alpha, 
\ \ \ \ \ \ \ \ \
k_{-1} = \alpha \exp (\Delta G/RT),
\label{2state_rates}
\end{eqnarray}
where $\alpha$ is sequence independent.  
Figure~\plainref{fig_2and3}(a) shows $\theta$ vs. $\Delta G$ for a fixed
concentration $c$ (see Eq.~(\plainref{2state_kin})) on a semilogarithmic
scale at five different hybridization times (dashed lines), where
the rates are chosen according to Eqs.~(\plainref{2state_rates}). The solid
line coincides with the equilibrium regime (limiting case $t \to \infty$).  The dashed
lines and the equilibrium isotherm merge for
small binding affinities. This indicates that weakly-bound sequences
equilibrate faster than more strongly bound ones.  This behavior is a
consequence of Eqs.~(\plainref{2state_tau}) and (\plainref{2state_rates}): The
relaxation time decreases for weak binding.
In the limiting case of strong binding ($\Delta G$ far from zero i.e. large $\tau$) and
short times, the expansion of Eq.~(\plainref{2state_kin}) yields $\theta (t) \approx \alpha c t$
which no longer depends on $\Delta G$.
\clearpage
 We refer to this as the {\it
dynamical saturation regime} which appears as a constant limiting behavior
in the $\theta$ vs. $\Delta G$ plot as shown in Fig.~\plainref{fig_2and3}(a).
Instead, we refer to {\it chemical saturation} as the limit $\theta
\to 1$, where all available probes are hybridized, and no further
hybridization is possible.  We emphasize that the existence of the
dynamical saturation limit is linked to the choice of the rates $k_1$,
$k_{-1}$ (see Eqs.~(\plainref{2state_rates})), where we assumed that $k_1$
is sequence independent.

\begin{figure}[t]
\includegraphics[width=8.5cm]{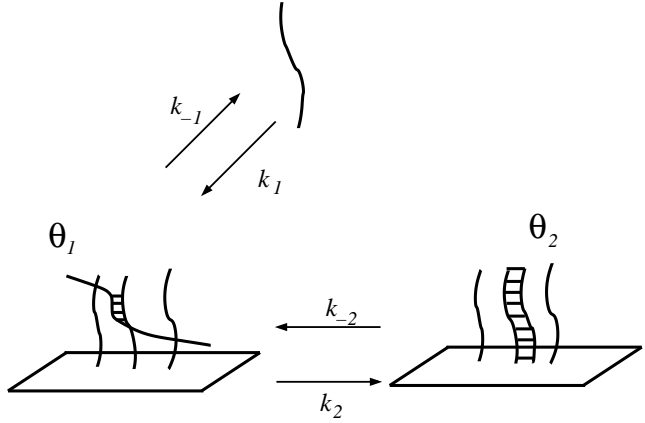}
\caption{Sketch of the three-state model for hybridization in DNA
microarrays. $\theta_1$ and $\theta_2$ are the fractions of partially
and fully hybridized strands respectively. 
This model is specified by the four rate constants.}
\label{fig3state}
\end{figure}

In view of the subsequent discussion of experimental data, we review
a recently introduced extension of the two-state model~\cite{hooy10}.
This extension includes an intermediate long-lived configuration of
partially hybridized molecules as sketched in Fig.~\plainref{fig3state}.
The presence of the intermediate state slows down hybridization to a
fully formed duplex due to the interaction of the
target molecule with multiple probes. 
There are now four rate constants
and the model is defined by two coupled linear equations
\begin{subequations}\label{dtheta}
\begin{eqnarray}
\frac{d\theta_{1}}{dt} &=& c k_1 (1-\theta_1 -\theta_2 ) +
k_{-2} \theta_{2} - (k_{-1} + k_2) \theta_1,
\label{dtheta1dt} \\
\frac{d\theta_2}{dt} &=& k_2 \theta_1 - k_{-2} \theta_2,
\label{dtheta2dt}
\end{eqnarray}
\end{subequations}
where $\theta_1$ and $\theta_2$ denote respectively the fractions of partially and
fully hybridized states. The ratios between the forward and backward
rates are fixed by thermodynamics
\begin{subequations}\label{thermo2}
\begin{eqnarray}
\frac{k_1}{k_{-1}} = e^{-\Delta G'/RT},
\end{eqnarray}
\begin{eqnarray}
\frac{k_2}{k_{-2}} = e^{-(\Delta G - \Delta G')/RT},
\end{eqnarray}
\end{subequations}
where $\Delta G$ and $\Delta G'$ are the hybridization free energies
of the fully and partially hybridized states, respectively
 (both are free energy differences with respect to the unhybridized state).
  In general, $\theta_1$ describes a distribution of partially
hybridized states where only a fraction of the target-nucleotides
is bound to a probe and the associated free-energy $\Delta G'$
can be considered as an effective $\Delta G$.  We expect the fully
hybridized state to be energetically more favorable than the partially
hybridized state,~i.e.~$\Delta G < \Delta G'$.  The forward reaction
rates $k_1=\alpha$ and $k_2=\omega$ are assumed to be sequence
independent while the reverse rates are fixed by the relations in
Eqs.~(\plainref{thermo2}). We then assume a monotonic link between
$\Delta G$ and $\Delta G'$.  Consider two sequences with different
total hybridization free energies where the first one has higher binding
affinity than the second $\Delta G_1 < \Delta G_2$ (Note that this may
refer to the same probe with one being a perfect match and the other
a mismatching sequence).  For the partially hybridized states, we then
expect $\Delta G_1' < \Delta G_2'$.  To establish a relationship between
the two free energies, we make the simple assumption
\begin{eqnarray}
\Delta G' \approx \gamma \Delta G,
\label{def_gamma}
\end{eqnarray}
where $\gamma <1 $ since $\Delta G < \Delta G'$.  With this input we
numerically solve Eqs.~(\plainref{dtheta1dt}) and~(\plainref{dtheta2dt}).
The solution can be expressed in terms of eigenvalues and eigenvectors of
a $2 \times 2$ matrix. Figure~\plainref{fig_2and3}(b) plots the solution
for the fraction of hybridized probes $\theta = \theta_1 + \theta_2$
for five different times (dashed lines). The solid line corresponds to
the equilibrium isotherm, which is identical to that of the two-state
model in Fig.~\plainref{fig_2and3}(a).  In addition, we find a regime of
dynamical saturation for very strongly bound sequences, which is analogous
to that of the two-state model.  The difference between the two models is
the emergence of a new nonequilibrium regime. There, the logarithm of
the fraction of hybridized probes varies linearly with $\Delta G$ with
a slope equal to $\gamma/RT$, where the parameter $\gamma$ is defined
in Eq.~(\plainref{def_gamma}).  We note that a regime characterized by a
slope smaller than expected from the equilibrium isotherm, is equivalent
to introducing an effective temperature $T_{\rm eff} = T_{\rm exp}/\gamma$
 higher than the experimental one.

\begin{table*}[t]
{\small
\begin{tabular}{l|l|l|l}
  & Agilent & Affymetrix & Codelink\\
\hline
Temperature (\textcelsius) & 55, 65 & 45    & 45 \\
Hybridization time (hours) & 17, 86 & 16    & 12 \\
Target length (nucleotides)& 25, 30 & 14-25 & 70 \\
Probe length (nucleotides) & 25, 30 & 25    & 30 \\
Number of different targets& 1      & 150   & 8  \\
Target concentration (picomolar) & $50$, $500$, $5000$ & 
		$1.4\cdot10^{-3}$-$1.4\cdot 10^3$ & $3000$
 \end{tabular}
}
\caption{Overview of experimental details of the three 
platforms analyzed in this paper.}
\label{tab:overview_expts}
\end{table*}

We conclude this section with a discussion of the dependence of $\theta$
on the target concentration. Figures~\plainref{fig_2and3}(a) and (b)
show $\theta$ at a fixed concentration with varying hybridization times.
Let us now consider a protocol in which the concentration is varied and
the time fixed.  In all cases far from the chemical saturation limit
$\theta \ll 1$, the data scale linearly with the concentration in the two
as well as the three-state model. This behavior is a consequence of the
form  of Eqs.~(\plainref{two_states_kin}) and~(\plainref{dtheta1dt}):
far from chemical saturation we approximate $c k_1 (1- \theta) \approx
c k_1$ in Eq.~(\plainref{two_states_kin}), and $c k_1 (1- \theta_1 -
\theta_2) \approx c k_1$ in Eq.~(\plainref{dtheta1dt}), such that the
concentration $c$ only enters
 in the source term of the differential equations in the two models. As a
result $\theta$ is proportional to $c$ far from chemical saturation. This
is illustrated in Fig.~\plainref{fig_2and3}(c) where the hybridization
time is fixed to $t=17h$. The concentration values range from $c=5pM$
to $c=500nM$ in increments of a factor $10$ (from bottom to top in the
main graph). Only at the highest concentration the system has reached
equilibrium after $t=17h$ (solid line), at all other concentrations the
system is not equilibrated. This is a consequence of a dependence of
the equilibration time on $c$ (see Eq.~(\plainref{2state_tau})
 for the two-state model).  The inset of Fig.~\plainref{fig_2and3}(c)
 shows the intensity divided by the concentration.  For the two lowest
 concentrations the
$\theta/c$ plots overlap for the whole range of $\Delta G$, while there
is no overlap if $\theta$ is close to $1$.

\section{Experiments}

In this Section we discuss the experimental results obtained by different
groups on three different microarray platforms. The details of the
experimental setups are summarized in Table~\plainref{tab:overview_expts}.
These experiments are performed on samples containing a controlled
number of different sequences in solution at some fixed concentrations.
Agilent arrays held one, Codelink 8 and Affymetrix 150 different 
target sequences.  In all cases the target hybridizes to a perfect
match probe and to a number of mismatching probes. The emphasis of the
analysis lies on the behavior of the signal intensities as a function of
hybridization free energies $\Delta G$. To estimate these free energies we
use the nearest-neighbor model \cite{bloo00}, according to which $\Delta
G$ can be written as a sum of parameters depending on the identity and
orientation of neighboring nucleotides. Nearest-neighbor parameters
from experiments on hybridizing strands in bulk solutions have been
intensively investigated in the past and are available in the literature
(see e.g. \cite{sant98}). As a first approximation, these parameters
could also be used for hybridization in microarrays. In controlled
experiments the parameters can also be extracted from microarray data,
see for instance Ref.~\cite{hooy09}. In the following $\Delta G_{\rm sol}$
refers to estimated free energies from experiments in bulk solution
 (data from~\cite{sant98}) and $\Delta G_{\rm \mu array}$ to estimated
free energies from microarray data.

\subsection{Custom Agilent arrays}

\begin{figure}[H]
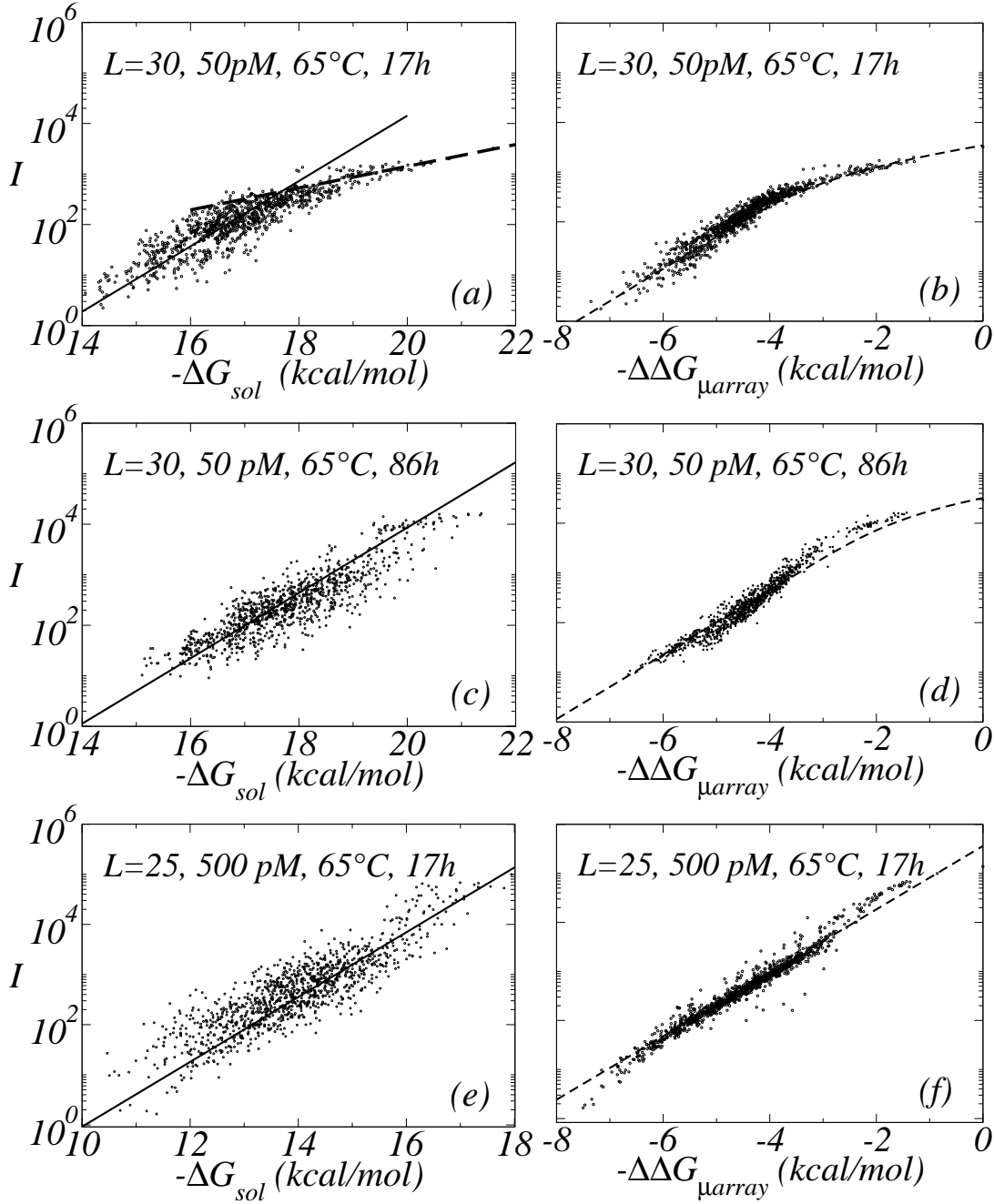

\hbox{
\includegraphics[width=0.465\columnwidth]{GraphNAR65C17hParamSol.eps} 
\includegraphics[width=0.4\columnwidth]{GraphNAR65C17hParamNAR.eps}
}
\hbox{
\includegraphics[width=0.465\columnwidth]{GraphNAR65C86hParamSol.eps} 
\includegraphics[width=0.4\columnwidth]{GraphNAR65C86hParamNAR.eps}
}
\hbox{
\includegraphics[width=0.465\columnwidth]{GraphOPT65C17hParamSol.eps}
\includegraphics[width=0.4\columnwidth]{GraphOPT65C17hParamOPT.eps} 
}

\caption{\textit{Left column:} $I$ vs. $\Delta G_{sol}$,
\textit{right column:} $I$ vs. $\Delta\Delta G_{\mu\rm{array}}$
obtained under different experimental conditions in an Agilent platform.
The full lines in (a,c,e) are the Langmuir isotherm with $T_{\rm exp}$,
as expected at equilibrium. The dashed line in (a) is the Langmuir
isotherm with $T_{\rm eff}=3T_{\rm exp}$ (intermediate regime). 
The dashed lines (b,d,f) are fitted using the three-state model with
$\gamma=1/3$, $k_1=5.10^3M^{-1}s^{-1}$, $k_2=1s^{-1}$, and we introduce
a shift of $7.5kcal/mol$ on the $\Delta G$ axis.  The amplitudes
are slightly adjusted to match the intensity scale in the different
experiments.  (a,b): Data collapse using $\Delta\Delta G_{\mu\rm{array}}$
of~\cite{hooy09} (\textit{right}) better than for parameters obtained in
solution (\textit{left}).  (c,d) Intermediate
regime almost vanished since equilibration is reached for nearly all sequences. The
effect is more pronounced for $\Delta\Delta G_{\mu\rm{array}}$ than for
$\Delta G_{sol}$.  (e,f) Faster equilibration
for shorter sequences: Compared to longer strands, the binding free energy
here is smaller, resulting in a shorter characteristic time. 
(a,b): Data from~\cite{hooy09}. 
(c,d): Data from~\cite{hooy10}.
(e,f): New experiment.}
\label{Fig:Jef&al1}
\end{figure}

We review some experimental results recently obtained by two of
us \cite{hooy09,hooy10} and present novel experimental data of
custom Agilent arrays. In this setup only one sequence is present in
solution at various concentrations. The microarray is then designed
to contain probes with either one, two or no mismatches with respect
to the target sequence in solution. In total there are about $10^3$
different probe sequences, replicated 15 times on the array (for the
details see Ref.~\cite{hooy09}).  The experimental data are shown in
Fig.~\plainref{Fig:Jef&al1}, which reproduces a pairwise comparison of the
logarithm of the same intensities vs. $\Delta G_{\rm sol}$ (left column)
and vs. $\Delta G_{\rm \mu array}$ (right column) for different target
lengths and hybridization times.
 Here, $\Delta G_{\rm sol}$ is calculated
based on nearest neighbor parameters by SantaLucia \cite{sant98} at 1M
[Na+] concentration. The hybridization buffer in the experiment is the
standard Agilent buffer which contains a buffering agent, monovalent
cation and a ionic surfactant. It is known that the salt concentration
influences the stability of the double helix and, thus also $\Delta
G$. However it usually affects all nearest-neighbor parameters by a
constant salt-dependent constribution \cite{sant98}. A change in salt
concentration (and presumably of other chemicals) would result in a
global shift of the horizontal axis in Fig.~\plainref{Fig:Jef&al1}. This,
however, does not influence our analysis. More details about the
effect of salt concentration and about electrostatic interactions 
are discussed in the Appendix 1.
The $\Delta G_{\rm \mu arrays}$ are obtained from a linear least-square fit
of the data, as explained in Ref.~\cite{hooy09}. The left column of
 Fig.~\plainref{Fig:Jef&al1} plots the data as function of $\Delta \Delta G_{\rm \mu array}=\Delta
G_{\mu\rm{array}}-\Delta G_{PM\mu\rm{array}}$, i.e. the subtraction from the
hybridization free energy of the perfect match probe, which cannot be
determined from the fitting procedure~\cite{hooy09}. The similarities
of the plots in both columns of Fig.~\plainref{Fig:Jef&al1}
show a correlation between $\Delta G_{\rm sol}$ and $\Delta G_{\rm
\mu array}$~\cite{hooy09}. However, when plotted as a function of $\Delta
\Delta G_{\rm \mu array}$ the data ``collapse" better into single master
curves, compared to the plots as a function of the nearest-neighbor model
parameters in solution which are more spread. 

Figures~\plainref{Fig:Jef&al1}(a,b) show the measured intensities for
a target strand of length $L=30$ hybridizing for $17h$, which is
the typical hybridization time of the standard protocol used in
biological experiments.  In the cases (c,d) the time was increased to
$86h$, while the cases (e,f) correspond to a shorter target sequence
($L=25$) hybridized for $17h$. All data are obtained at a temperature
of $T=65$\textcelsius. Experiments were also repeated under the same
conditions at different target concentrations; a change in
global concentration leads to an overall multiplicative factor in the
measured intensities as explained in the Appendix 2.

The solid lines in Fig.~\plainref{Fig:Jef&al1}(a,c,e) correspond to the slope
expected in the case of thermodynamic equilibrium ($1/RT_{\rm exp}$), 
the dashed line in Fig.~\plainref{Fig:Jef&al1}(a) corresponds to the Langmuir
 isotherm with an effective temperature $T_{\rm eff}=3T_{\rm exp}$, 
whereas the dashed lines in (b,d,f) are obtained from the solution of the
three-state model. The measured intensity is assumed to be proportional
to the fraction of hybridized probes $I= A \theta$. The scale factor $A$
is slightly adjusted in the dashed lines of Fig.~\plainref{Fig:Jef&al1}(b,d,f)
to match for a global scale difference in intensities in the different
experiments. Apart from the rescaling the dashed lines are obtained by using
the same values for the adjustable parameters in the three-state model
(the values are given in the caption).  When increasing the hybridization time
(from (b) to (d)) a larger fraction of the data tend to align along the
expected equilibrium isotherm. Full equilibration is reached after $17h$
when using a shorted target length (f), consistent with the three-state
 model prediction: shorter sequences have lower binding affinity
(lower $-\Delta G$), and are therefore expected to equilibrate faster.

\begin{figure}[H]
\includegraphics[width=0.75\columnwidth]{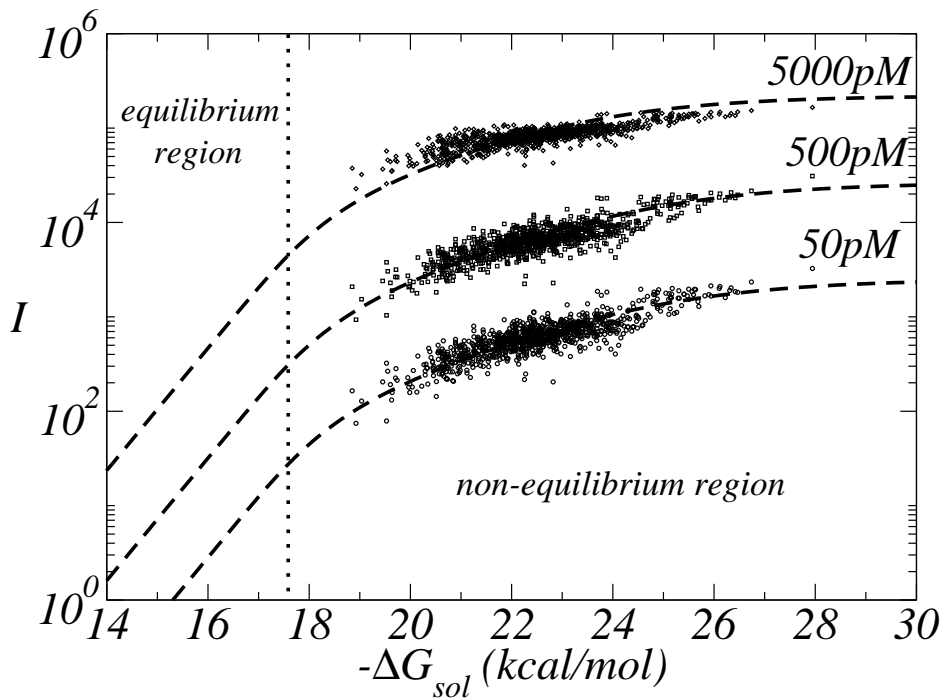}
\caption{$I$ vs. $\Delta G$ for $c=\{50,500,5000\}pM$, $L_{probe}=L_{target}=30$, $T_{exp}=55$\textcelsius\,
 and a hybridization time of $17h$. Lowering
the temperature shifts the free energies to higher $-\Delta G$'s
compared to Fig.~\plainref{Fig:Jef&al1}(a). 
The dotted lines are a fit from the three-state model
 using the same parameters as Fig.~\plainref{Fig:Jef&al1}.
 Note that the shape for the largest concentration is
flattened due to the saturation of the scanner around $10^5$.}
\label{Fig:Jef&al2}
\end{figure}

Figure~\plainref{Fig:Jef&al2} shows experimental data for
$L=30$ and a hybridization time of $17h$ at a temperature of
$T=55$\textcelsius. Lowering the temperature leads to higher binding
affinities (higher $-\Delta G$), which causes an increase in the
relaxation time according to the three-state model. We therefore
expect that these data would be even further away from thermodynamic
equilibrium. This is indeed experimentally confirmed. In the present case
none of the data, even for the lowest intensities attains the limiting
behavior $I \propto e^{-\Delta G/RT}$, which suggests a breakdown of
thermal equilibrium.  Note that we did not plot the data as function of
$\Delta \Delta G_{\rm \mu array}$ since it is only possible to extract
these parameters if equilibrium data are available, which is not the case
at $T=55$\textcelsius. The dashed lines in Fig.~\plainref{Fig:Jef&al2}
are fits of the three-state model, with the same fitting parameters
compared to the case $T=65$\textcelsius\ (see figure caption).  The data
at the highest affinities tend to bend towards a flat asymptotic limiting
behavior, which suggests that they approach the dynamical saturation
limit, discussed in the previous section. For the sake of completeness,
we mention that the signal of the highest concentration is even more
flattened due to the saturation of the scanner around $10^5$.  Note also
that there is a linear scaling with the concentration and this is verified
in a range of two orders of magnitude in $c$ (from $50$ to $5000pM$).

\subsection{Custom Affymetrix Microarray}

We discuss next the experiments by Suzuki et al~\cite{suzu07}.  In these
experiments the arrays were custom-designed and synthesized on the
Maskless Array Synthesizer platform with the Affymetrix NimbleExpress
program. The target length was $L=25$ nucleotides throughout the
experiments. $150$ artificial target sequences with a CG-content ranging
from $32$\% to $72$\% were used such as to match typical distribution
of probes of the Affymetrix E. Coli array. The probe lengths varied
from $L=14$ to $L=25$ nucleotides. The probes were shortened from the
5'-end. This implies that hybridization between probes with $L<25$
nucleotides and targets produced duplexes with dangling ends afar from
the microarray surface.  For each target sequence three corresponding
PM sequences and all possible mismatch combinations were present on the
surface,~i.e.~three mismatches per position along the strand.  In order
to avoid terminal effects (it is known that mismatches close to the helix
ends have different stability properties compared to mismatches in the
center of the helix \cite{sant04}), our analysis considers only those
mismatches which are at least three nucleotides spaced from the ends.
The concentration of each target sequence ranged from $1.4fM$ to $1.4nM$
incrementing in steps of factor ten; a concentration of a single target
of $1.4nM$ corresponds to a total concentration of $1.4 \times 150 =
210nM$.  The hybridization time was $16h$ at $T=45$\textcelsius.
The data cover a broad range of concentrations and $\Delta
G$, compared to the setup discussed in the previous section.

\begin{figure}[H]
\includegraphics[width=0.75\columnwidth]{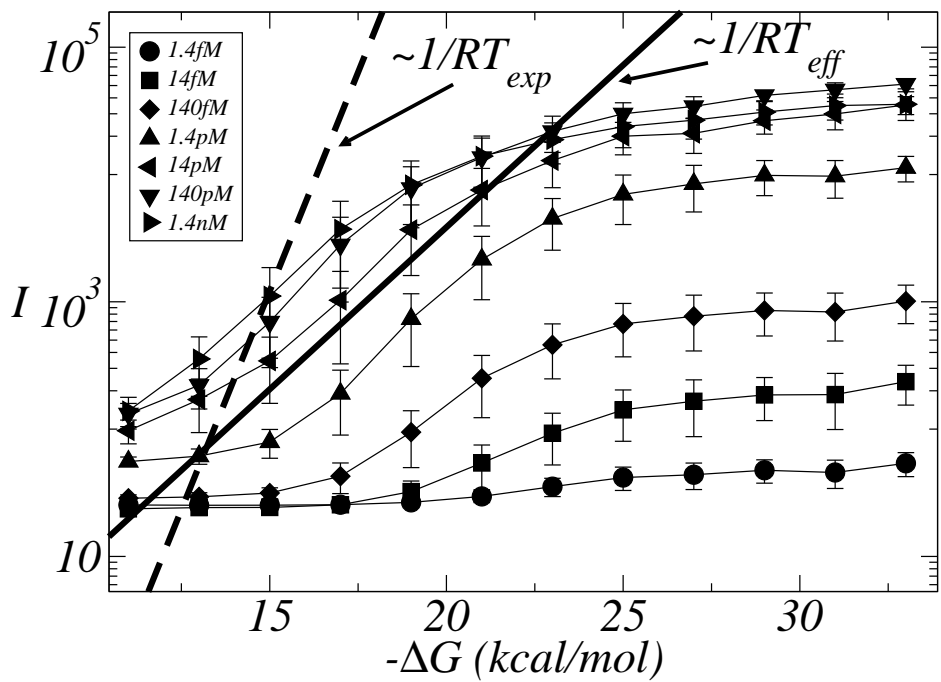}
\caption{$I$ vs. $\Delta G$ curves for dataset 1 obtained by Suzuki
et al~\cite{suzu07}.  Binding free energies $\Delta G$ determined by
means of nearest neighbor model using the values obtained in solution~\cite{sant98}.
 Saturation occurs at
different values for different concentrations.  The dotted and solid
lines correspond to the Langmuir isotherm using respectively  $T_{\rm
exp}=318K$ and $T_{\rm eff}=850K\approx2.7T_{\rm exp}$.}
\label{average_jap}
\end{figure}

Figure~\plainref{average_jap} shows a $\log I$ vs. $\Delta G_{\rm sol}$ plot of
the data for different values of the total concentration.
  A large number of data is available for each experiment ($78300$ data points including
the PM's and single mismatches of the $150$ different targets). To help
the visualization of the data were binned in intervals $[\Delta
G-\delta/2,\Delta G+\delta/2]$ with $\delta = 2kcal/mol$. We then
calculate the median value of the intensities within one $\Delta G$
interval of all probe sizes. The error bars in the graphs provide an
estimate of the typical spreading of the data using the mean absolute deviation.
For very weak binding ($-\Delta G < 10kcal/mol$), the data
of Fig.~\plainref{average_jap} tend to a constant background value. The
background level increases as the concentration is increased, due to
aspecific binding, as expected.  The slope of the dashed line in Fig.~\plainref{average_jap}
is $1/RT_{\rm exp}$. The solid line (which provides a good extrapolation of
the low $-\Delta G$ data) corresponds to an ``effective" temperature
of $T_{\rm eff} = 850K$ (a value close to the one found in the analysis
of the experiments in the Agilent platform). Thus, these experimental
data cannot be reconciled with equilibrium thermodynamics. In
addition, we note that for strong binding affinities
 the data reach a saturation value. This behavior would be consistent to
the dynamical saturation predicted by the kinetic models discussed in 
the section describing the two- and three-state models.

\begin{figure}[H]
\includegraphics[width=0.75\columnwidth]{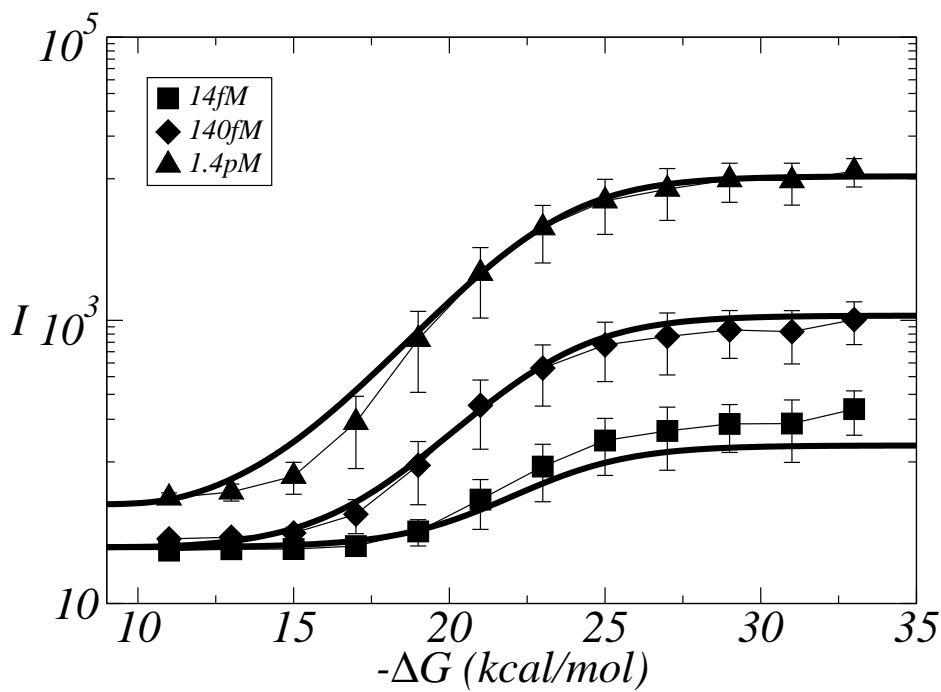}
\caption{For three intermediate concentrations, the experimental data are
well fit by the three-state model (solid thick lines).  The parameters
are: $\gamma=0.374$, $k_1=10^5 {s}^{-1}{M}^{-1}$ and  $k_2=1{s}^{-1}$.
 The value are scaled by a factor $A=2^{18}$ in order to match
the experimental data.  We have adjusted the value of the background $I_0$
for each concentration.}
\label{average_jap3state}
\end{figure}

In Fig.~\plainref{average_jap3state} we plot the three-state model isotherm
(thick solides lines) corresponding to an hybridization time of $16h$ as in the
experiment and using the values for the parameters $k_1$, $k_2$
and $\gamma$ given in the caption.
 The model fits well the data within two orders of magnitude in
concentration. To fit the data in the low $-\Delta G$ limit we added
a background value ($I_0$)~: $I=I_0 + A \theta$ (where $\theta = \theta_1 + \theta_2$ is the total fraction of bound
probes as calculated from the three-state model). Higher concentrations
are also well fitted by the three-state model, provided the adjustment of the
global fluorescence intensity level by rescaling the prefactor $A$.
This rescaling is necessary since at high concentrations the intensities
in Fig.~\plainref{average_jap} no longer scale linearly with the concentration,
differently from what observed in all the experiments on Agilent
arrays (see Fig.~\plainref{Fig:Jef&al2} and Refs.~\cite{hooy09,hooy10}).
This behavior is to be expected close to chemical saturation $\theta
\to 1$, but not in the region $15kcal/mol < -\Delta G < 20kcal/mol$, where the measured
intensities are very far from saturation. It suggests that other effects
may play a role such as e.g. interactions (partial hybridization) between target
molecules in solution. The experimental data of Fig.~\plainref{average_jap} have also been
analyzed in the recent literature~\cite{ono08_sh,burd10} and were
fitted with a model taking target depletion into account. Depletion
arises when a relevant fraction of the target molecules in solution
hybridizes to the microarray, so that during the experiment the
concentration in solution decreases.  Depletion would explain why the
intensities at high binding affinity reach a saturation value below
the chemical saturation~\cite{ono08_sh,burd10}, without the need of
invoking a nonequilibrium effect. However the depletion models of
Refs.~\cite{ono08_sh,burd10} did not addressed the differences between
the experimental data and equilibrium thermodynamics in the low binding
limit ($-\Delta G < 20kcal/mol$ in Fig.~\plainref{average_jap}), whereas
the nonequilibrium scenario developed within the three-state model
could explain the experimental data in both regimes.
 It remains
to be proven that the observed behavior is due to breakdown of thermal
equilibrium. One possibility would be to perform hybridization experiments
at higher temperatures. According to the three-state model higher
temperatures (recall that the experiments of Fig.~\plainref{average_jap}
are at $T=45$\textcelsius) would favor equilibrium due to a decrease
in $-\Delta G$. This is indeed observed in Agilent arrays (see
Figs.~\plainref{Fig:Jef&al1},~\plainref{Fig:Jef&al2}). We recall that
other effects could also explain why the slope of the $I$ vs. $\Delta G$
data is smaller than what expected from the Langmuir model. A recent study
\cite{nais09} showed that synthesis errors (i.e. the synthesis of
incorrect nucleotide) along the probe may also influence this slope.
At present, further experiments are necessary to clarify the origin
the discrepancy between the Affymetrix data and the equilibrium Langmuir 
model.


Finally, we used the experimental data to perform a linear least square fit
to obtain estimates of the nearest neighbor parameters.  The interval range
 was chosen such that $\log I\propto\Delta G$,
far from any saturation behavior. We
chose the two largest concentrations $c=140pM$ and $c=1.4nM$,
focusing on the hybridization of size $L=20$ with a range of $\Delta
G$ between $13.5kcal/mol$ and $19kcal/mol$. In this range $\log
I$ is approximately linear as a function of $\Delta G$ as seen
in Fig.~\plainref{average_jap}.  The fitting procedure is similar to the one 
described in Ref.~\cite{hooy09}, and it is obtained through singular
value decomposition from the minimization of a quadratic cost function.
The procedure provides $58$ nearest neighbor parameters, of which
$10$ refer to perfect match parameters and $48$ to single mismatch parameters.
 Due to degeneracies~\cite{gray97a}, some of these parameters are not unique.
 In order to compare them with the corresponding parameters in solution,
we consider the combination:
\begin{eqnarray}
  \Delta\Delta G
  \begin{pmatrix}
  T&\underline{A}&C\\
  A&\underline{C}&G
  \end{pmatrix}
=
  \Delta G
  \begin{pmatrix}
  T&\underline{A}&C\\
  A&\underline{C}&G
  \end{pmatrix}
-
  \Delta G
  \begin{pmatrix}
  T&A&C\\
  A&T&G
  \end{pmatrix}
,
\nonumber
\end{eqnarray}
which is instead unique~\cite{hooy09}.
\begin{figure}[h!]
\begin{center}
\includegraphics[width=0.75\columnwidth]{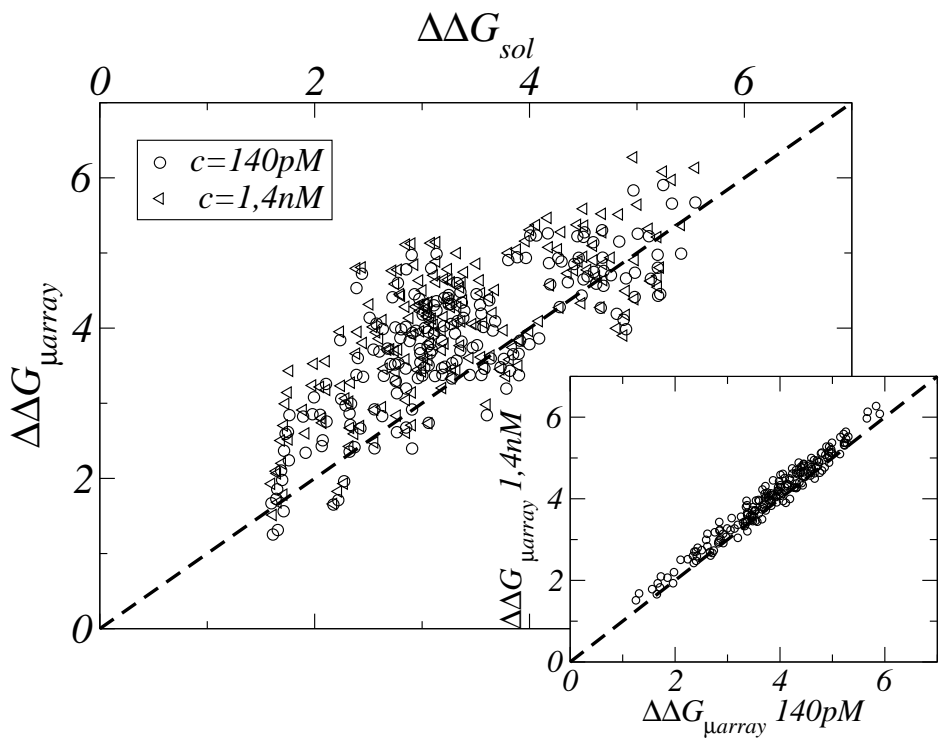}
\caption{Main graph: $\Delta\Delta G_{\mu array}$ vs. $\Delta\Delta G_{sol}$. The data of Suzuki et al~\cite{suzu07}
are fitted (probe size $L=20$) using the linear least-square method in the interval $-\Delta
G$ between $13.5$ and $19kcal/mol$ for two concentrations $140pM$
and $1.4 nM$. We have used the effective temperatures $T_{eff}=750K$
for $140 pM$ and $T_{eff}=960K$ for $1.4 nM$.  
The Pearson coefficient between the two axes is $0.784$ for $140pM$ and
$0.735$ for $1.4nM$, showing a moderate correlation.  Inset:
the results for both concentrations are plotted with respect to each
other in order to check the cross-correlation
 (Pearson coefficient between the two concentrations: $0.983$).}
\label{Fig:Japanese4}
\end{center}
\end{figure}
Figure~\plainref{Fig:Japanese4} shows a comparison between the parameters
obtained from a fit of the microarray data compared to those obtained
from hybridization/melting experiments in Ref.~\cite{sant98}.
 The Pearson correlation coefficients are $0.784$
for $140 pM$ and $0.735$ for $1.4 nM$, indicating a moderate correlation
between the two sets. The inset shows a plot between parameters obtained
from the $140 pM$ experiment vs. those of the $1.4 nM$ experiment.
There is a strong correlation between these two sets (Pearson coefficient
$0.983$) confirming the robustness of the extraction procedure of the parameters. 
However, note that this last analysis has to be considered with care. 
The validity of the estimation of the parameters in a regime where the use of
 an effective temperature is needed (and possibly a regime out-of-equilibrium) is still an open question. 

\subsection{Codelink activated slides}

Last, we present another set of experiments on custom Codelink
activated slides by Weckx et al.~\cite{weck07}. In these experiments the
target solution contained 8 different sequences. Note that only four were
considered in the analysis (since the other four are richer in AT and bind
very weakly to their probes, having a signal just above the the background
level).  The target concentration, temperature and hybridization times
were $c=3000 pM$, $T=45$\textcelsius, and $12h$ respectively.  The target
is a $70$-mer, with a stretch of $30$ nucleotides complementary to the
probes and $40$ nucleotide spacer terminating with a fluorophore. The
spacer keeps the fluorophore away from the hybridization area and from
the probe layer.  Twelve probe sequences were associated to each target:
one perfect match, $3$ carry one mismatch, $7$ carry two mismatches and
one has $3$ mismatches. This sums up to $48$ data points. Each probe was
 replicated $12$ times on the array, the analyzed signal is the median
over the replicates.

\begin{figure}[H]
\begin{center}
\includegraphics[width=0.75\columnwidth]{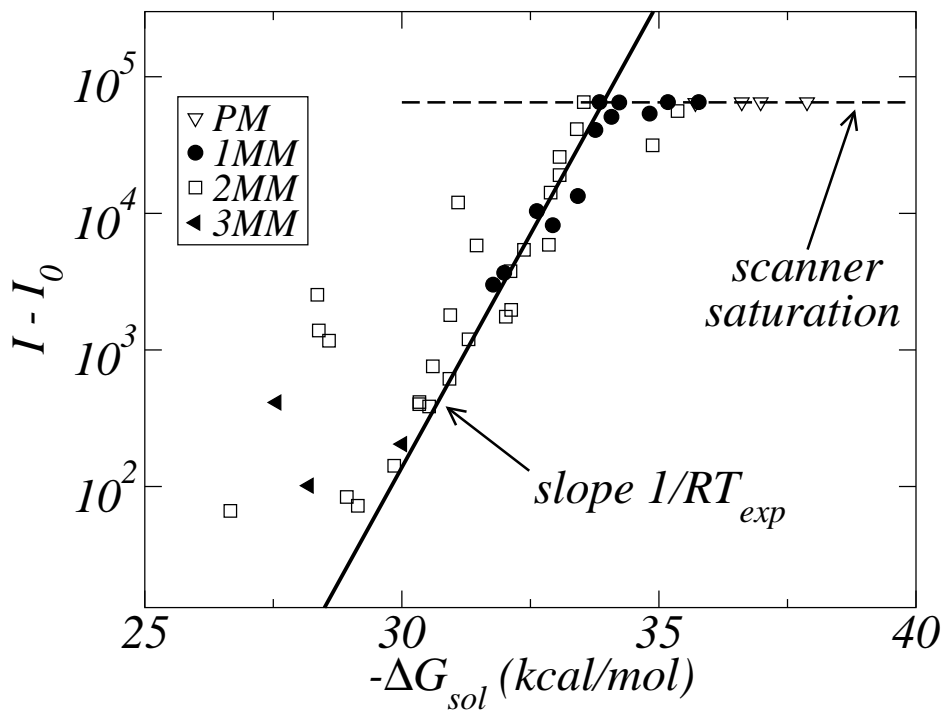}
\caption{Background subtracted intensities 
vs. hybridization free energies in solution for the experiments of 
Weckx et al.~\cite{weck07} on custom Codelink activated slides. 
The data are generally in agreement with equilibrium thermodynamics
as seen from the alignement along a line with slope
$1/RT_{\rm exp}$. The experimental temperature is $T=45^\circ$C
 for a hybridization time of $12h$.}
\label{Fig_stefan}
\end{center}
\end{figure}

The experimental results are shown in Fig.~\plainref{Fig_stefan}.
The data are plotted as a function of $\Delta G_{\rm sol}$ calculated
from the nearest-neighbor parameters for hybridization in solution
by~\cite{peyr99}. The horizontal line corresponds to the maximum intensity
detectable from the scanner (optical saturation). The steep line has a
slope equal to $1/RT_{\rm exp}$. Most probes with low $\Delta G$ are
distributed in the close vicinity of the $1/RT_{\rm exp}$ line. This
suggests that the system had attained thermodynamic equilibrium.
Although the number of data point is insufficient to extract free energy
parameters, the alignment of the majority of the data points along a
line with slope $1/RT_{\rm exp}$ suggests a good degree of correlation
with the hybridization free energy parameters measured in aqueous
solution~\cite{peyr99}.
In Codelink activated slides the probes are not directly bound to the
solid surface, but to a three dimensional polymeric coating which shifts
them further away from the surface.
 This is to reduce the steric hindrance and to facilitate hybridization.
 The reduced surface probe density will also reduce electrostatic
interactions from the negatively charged DNA probes.
 This is probably why these systems attain thermal equilibrium faster than those presented above.

\section{Conclusion}

Although several papers have discussed the application of physico-chemical 
models to describe microarray data, so far most works focused on equilibrium
thermodynamics~\cite{held03,heks03,zhan03,haga04,bind05b,carl06,fish07_sh,weck07,zhan07,burd08,nais08a}.
The tacit assumption behind the use of these models is that hybridization
 in typical microarray experiments is sufficiently long
(usually $15-17h$), so that thermal equilibrium can be considered
to be attained. However, in standard microarrays, which are based on
fluorescence detection,  it is not possible to perform real time
measurements of the hybridization signal. then, it is difficult to test this hypothesis directly
from experimental data.

In this paper we discussed how the attainment of equilibrium can
be checked in experiments. We analyzed experiments from three
different microarray platforms where the number of mismatches per probe varied from 0 to 3. This is an experimental
setup which can offer very interesting insights about the physical
properties of hybridization. First of all, given a sufficiently
large number of mismatches, one can cover a broad range of intensities
in an experiment even with a single hybridizing sequence, avoiding
thus interactions between different targets in solution. Second, the
attainment of thermodynamic equilibrium can be tested from an analysis
of the slope of the experimental data in a $\log I$ vs. $-\Delta G$
plot. We expect a slope equal to $1/RT_{\rm exp}$ for equilibrium
data, whereas a deviation to a different slope can be taken as an
indication of the breakdown of equilibrium (sufficiently far from
chemical saturation).  We note also that in high density arrays used in
biological experiments, the slope of $\log I$ vs. $-\Delta G$ does not
usually match the experimental temperature \cite{held03,carl06}, which
could be a signature of nonequilibrium behavior, although different
explanations exist \cite{nais09}.

To explain the observed experimental results we discussed a three-state
model of hybridization. This is an intermediate between the unbound
state and the fully hybridized state. For finite hybridization times,
the three-state model predicts that sufficiently weakly bound targets
are in equilibrium, while nonequilibrium effects are expected to
occur for higher binding affinities. This feature is generally observed
in experiments.  In the limit of strong binding affinities the model
predicts the existence of a dynamical saturation limit, in which the
hybridization signal becomes independent on $\Delta G$. These features
are also observed in experiments on different platforms.


We expect that the absence of thermal equilibrium has important practical
consequences for the functioning of the microarrays also in biological
experiments with complex mixture of target sequences from mRNAs extracts.
%
In that case different target sequences ``compete" for hybridization
to the same probe. Let us consider the case of two competing targets,
one perfect matching to the probe and one carrying mismatches. The
hybridization of the probe with the latter produces a so-called
cross-hybridization signal. Assuming that the concentration of the
targets is low, the following
model is expected to hold:
\begin{eqnarray}
\theta = c f(\Delta G) + c' f(\Delta G')
\end{eqnarray}
where $c$ is the concentration of the perfect matching target and
$\Delta G$ its hybridization free energy; $c'$ and $\Delta G'$ refer
to the mismatched target. The function $f()$ describes the isotherm.
In thermal equilibrium $f(x) = e^{-x/RT_{\rm exp}}$.  A high specificity,
which is the desired working condition, corresponds to a 
perfect match signal dominating over the cross-hybridization contribution.
This is obtained when the ratio $f(\Delta G)/f(\Delta G') \gg
1$, being $c$ and $c'$ fixed in an experiment.  This ratio is maximal
in thermodynamic equilibrium. If, for instance, $\Delta G'-\Delta G=
3 kcal/mol$ (typically the free energy difference of one mismatch) at
equilibrium one has $f(\Delta G)/f(\Delta G') = \exp((\Delta G'-\Delta
G)/RT_{\exp}) \approx 90$. In the nonequilibrium regime characterized
by an effective temperature $T_{\rm eff}=3T_{\rm exp}$ one has $f(\Delta
G)/f(\Delta G') = \exp((\Delta G'-\Delta G)/RT_{\rm eff}) \approx 4$.
Thus the thermodynamic equilibrium corresponds to the highest specificity.

\begin{acknowledgement}

We are grateful to S. Weckx for providing us the data for
Fig.~\plainref{Fig_stefan}.  We thank Karen Hollanders (VITO) for help
with the experiments.  We acknowledge financial support from Research
Foundation-Flanders (FWO) Grant No. G.0311.08 and from KULeuven Grant
No. OT/07/034A.
\end{acknowledgement}

\section*{Appendix 1: The effect of salt concentration}

Isotherms different from the Langmuir model were discussed in the
microarray literatures of the last decade \cite{vain02}. Particular
attention was devoted to the the effect of electrostatic interactions,
which arise during hybridization between a target molecule and a dense
probe layer. As DNA is negatively charged, an additional electrostatic
repulsive force may arise when a target molecule approaches the probes 
at the microarray surface.

In the context of a mean-field approximation Vainrub and Pettit
\cite{vain02}, Halperin Buhot and Zhulina \cite{halp04} derived the
following equilibrium model
\begin{eqnarray}
\frac{\theta_{\rm eq}}{c (1-\theta_{\rm eq})} = 
e^{-\Delta G/RT -\Gamma(1+\theta_{\rm eq})/RT} 
\label{electrostatic}
\end{eqnarray}
where, as in the main text, $\theta_{\rm eq}$ is the fraction of
hybridized probes, $c$ the target concentration, $\Delta G$ the
hybridization free energy between a single probe and a single target.
The term $\Gamma (1+\theta_{\rm eq})/RT$ accounts for the electrostatic
repulsion, where $\Gamma$ is a constant independant of the sequence.
 Note that if $\Gamma =0$ one recovers from Eq.(\plainref{electrostatic})
the Langmuir isotherm of Eq.(\plainref{2state_eq}).  At high salt concentrations,
electrostatic interactions are screened. Equation (\plainref{electrostatic}) shows
that at larger coverages ($\theta_{\rm eq}$ large) the electrostatic
effects increase, reducing target-probe binding affinity.

The experiments discussed in this paper corresponds to the low
concentration limit, as the measured intensities in all cases results
proportional to the global target concentration $c$.  In the limit $c
\to 0$, Eq.(\plainref{electrostatic}) becomes
\begin{eqnarray}
\theta_{\rm eq} = c e^{-\Delta G/RT -\Gamma/RT} + \ldots
\label{electrostatic_lowc}
\end{eqnarray}
where the dots indicate higher orders in $c$.  Compared to the limit
obtained from the Langmuir model, the electrostatic effects provides an
additional contribution to the hybridization free energy ($\Gamma$).
In the mean-field model discussed in the literature, $\Gamma$
is proportional to the charge density of the unhybridized layer,
to its thickness and to the length of the target DNA. It is however
independent from the sequence composition.  $\Gamma$ is therefore the
same for hybridization to a perfect matching probe or to a probe with
one or two mismatches. In conclusion, in the low concentration limit,
 according to the above model, electrostatic effects can cause a
 uniform shift of the free energy scale,
compared to hybridization free energies in solution.


\section*{Appendix 2: The concentration scaling}

\begin{figure}[H]
\begin{center}
\includegraphics[width=0.75\columnwidth]{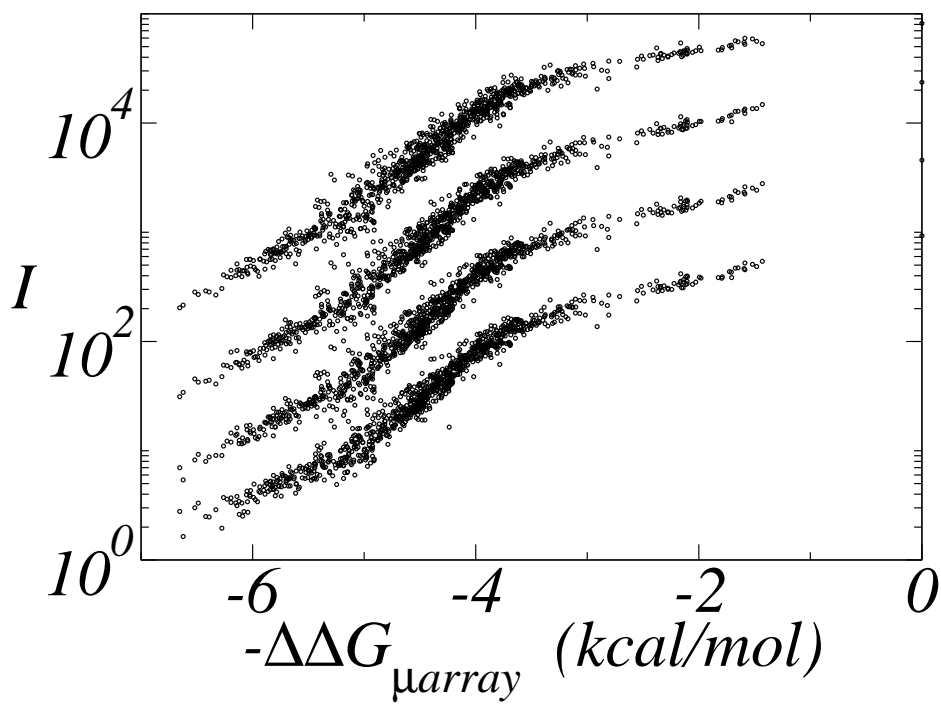}
\caption{Data obtained with Agilent platforms for a target and probe length of 30 nucleotides
 at $65^\circ$C with various concentrations $c=\{2,10,50,250\}pM$ from bottom to top (data from~\cite{hooy10}). 
The linear scaling in concentration for the different curves means that the kink does not correspond to
 the approach toward chemical saturation where the linearity does not hold anymore.}
\label{Fig_Scal_c}
\end{center}
\end{figure}

As discussed throughout the manuscript all experiments shown are
performed in a regime far from the chemical saturation limit which
corresponds to $\theta \to 1$. In the limit $\theta \ll 1$, $\theta$ and
thus the measured intensity, is proportional to the target concentration
$c$. We show this explicitly here for one set of experiments on Custom
Agilent arrays. Figure \plainref{Fig_Scal_c} shows the same data
as~\plainref{Fig:Jef&al1} $(a,b)$ ($L_{probe}=L_{target}=30$, $T=65^\circ$C)
for different concentrations $c=\{2,10,50,250\}pM$ from bottom to top.
The different curves clearly display an overall linear scaling
relationship with respect to the concentration, both in the weak
(equilibrium) and strong (nonequilibrium) binding regimes. The fivefold
increase in the concentration correspond to a fivefold increase of the
fluorescence intensity.

\end{document}